\documentclass[english]{article}
\usepackage[T1]{fontenc}
\usepackage[latin9]{inputenc}
\usepackage{textcomp}
\usepackage{amsmath}
\usepackage{amssymb}
\usepackage{graphicx}

\makeatletter

\newcommand{\lyxdot}{.}

\makeatother

\usepackage{babel}
\begin{document}
\title{Some fractal thoughts about\\
the COVID-19 infection outbreak}
\author{Massimo Materassi\thanks{Institute for Complex Systems of the National Research Council CNR-ISC,
via Madonna del Piano 10, 50019, Sesto Fiorentino (Florence), Italy,
e-mail: massimo.materassi@isc.cnr.it, massimomaterassi27@gmail.com,
web: www.materassiphysics.com}}
\maketitle
\begin{abstract}
Some ideas are presented about the physical motivation of the apparent
capacity of generalized logistic equations to describe the outbreak
of the COVID-19 infection, and in general of quite many other epidemics.
The main focuses here are: the complex, possibly fractal, structure
of the locus describing the ``contagion event set''; what can be
learnt from the models of trophic webs with ``herd behaviour''.
\end{abstract}

\section{Introduction\label{sec:Introduction}}

In these days of worldwide mourning for the human tragedy due to the
COVID-19 pandemic, while experiencing the heavy lockdown, in the fear
of a possible forthcoming economical and social crisis, I think every
scientist is thinking about how to be of help, in terms of her/his
ideas and technical culture. Besides humbly admitting this to be,
first of all, the work of doctors, nurses, virologists, biologists
and epidemiologists, not to mention the engineers who design medical
and life-saving devices, each of us wonders what technical tools among
ours might either help the fight, or at least teach something about
how the pandemic appears to work. Speaking for myself, I am doing
some job on the biomathematics of epidemics partly for scientific
curiosity and intellectual challenge, and partly (I'm sure) just to
do something in contact with what is happening out there, be it the
epical tragedy or the difficult clash against the virus. The antidote
my research is looking for is, at least, the one against the feeling
of personal powerlessness.

I felt the need to publish these notes when I went through the paper
\cite{Cinesi.Svizzeri}, in which the authors perform a deep investigation
about the behaviour with time of the ``total number $C\left(t\right)$
of people infected so far'', for the Chinese Province of Hubei, the
other Chinese Provinces, and few other countries undergoing the outbreak
of COVID-19, i.e. South Korea, Japan, Iran and Italy. The fact attracting
my attention, among the various results reported, was the apparent
good performance, in fitting the contagion data, of the law $C\left(t\right)$
solving the Generalized Richards Model (GRM) \cite{GRM}; this is
a modification of the logistic equation
\begin{equation}
\dot{C}=rC\left(1-\frac{C}{K}\right),\label{eq:logistica}
\end{equation}
in which the population variable appears with powers different from
$1$, typically between zero and $1$:
\begin{equation}
\dot{C}=rC^{p}\left[1-\left(\frac{C}{K}\right)^{\alpha}\right],\ p,\alpha\in\left[0,1\right],\ \alpha\ge p\label{eq:Sornette}
\end{equation}
(despite they are omonimous, the coefficients $r$ in (\ref{eq:logistica})
and (\ref{eq:Sornette}) are not the same quantity: in particular,
in (\ref{eq:logistica}) one has $\left[r\right]=t^{-1}$, while in
(\ref{eq:Sornette}) one has $\left[r\right]=C^{1-p}t^{-1}$).

The Ordinary Differential Equation (ODE) (\ref{eq:Sornette}) has
been largely and succesfully employed in epidemiology, so there is
no real surprise in its applicability to the COVID-19 outbreak. However,
as theorists usually do, I wonder which ``first principle meaning''
should be given to those powers $p$ and $\alpha$ in (\ref{eq:Sornette}).
In the literature, those parameters appear to be mostly guessed/adapted
empirically, in order to fit the experimental curve $C\left(t\right)$
a posteriori, once the epidemic is completely developed. Some study
noticeably put the $\left(p,\alpha\right)$ values in relationship
with the ``microscopic contagion dynamics'' \cite{Ebola.vs.HIV}:
it is shown that these constants depend on the \emph{geography of
contagion} (whether the epidemic develops in regions with or without
clusters of population and communications, like towns of various size,
or not) and on the \emph{sociology of contagion} (in the case of HIV,
for instance, whether this takes place via sexual intercourse, or
via needle sharing); these ``microscopic conditions'' appears to
be what gives the \emph{contagion networks} a different topology.
Recently, the specific study \cite{ZiffZiff} on the COVID-19 outbreak
argued how the scale-free complex clusterization of contagion events
could motivate a fractional kinetics for $C\left(t\right)$; I am
myself involved, together with Giuseppe Consolini, in a study about
fractional ODEs possibly solved by the COVID-19 $C\left(t\right)$
best fit in Italy \cite{Consolini.Materassi.COVID-19}.

All in all, it is clear that deviations of $C\left(t\right)$ from
the pure logistic ODE (\ref{eq:logistica}) points towards the departure
of the real laws governing the infection spread from the hypotheses
one has to assume to think a population $C$ satisfies (\ref{eq:logistica}).
One of the most important assumptions of the logistic equation, and
of all ODEs regulating the kinetics of population growth, is the so
called \emph{mass hypothesis} \cite{mass.hypothesis,Fanelli.Piazza.Covid}:
one assumes that the mixing of individuals is such, that all the microscopic
actions represented in the ODE are taken (on the average) by all the
elements of that population. This, roughly, gives rise to integer
powers of population variables in the ODEs. In a certain sense, one
might expect equation (\ref{eq:Sornette}) to be a consequence of
\emph{mass hypothesis violation} by the statistics of the microscopic
actions (\emph{contagion events}), of which it represents a mean field
description.

Why should that take place?

The answer I suggest here in inspired by the \emph{Trophic Web Theory}
(TWT), where the dynamics of interacting populations in ecosystems
are represented via coupled ODEs, in which a form of segregation may
take place, namely \emph{herd behaviour}: this assumption, that represents
a topological correction to the mass hypothesis, leads the population
variables to appear in their ODEs with real, possibly non-integer,
powers, exactly as in the GRM that the authors of \cite{Cinesi.Svizzeri}
claim to fit well the curves of the COVID-19 epidemic.

Here I present a personal vision on how to interpret the powers $p$
and $\alpha$ in (\ref{eq:Sornette}), according to the idea of herd
behaviour in TWT: my suggestion is that the ``abstract geometrical
locus of contagion events'' may have a \emph{fractal dimension},
under the assumption that the network of human contacts giving contagion
may be a scale-free network, with clusters of links on many scales,
just reflecting those geographical and sociological aspects evoked
by the authors of \cite{Ebola.vs.HIV}.

The paper is organized as follows.

In § \ref{sec:Logistic-and-generalized} the logistic ODE is recalled,
and its non-integer power generalizations are presented. In § \ref{sec:Herd-behaviour-and-fractals}
the concept of herd behaviour in TWT is introduced, with some examples
from ecology. Then, its interpretation of the parameters $p$ and
$\alpha$ is presented.

Conclusions, and possible developments of these ideas, are given in
§ \ref{sec:Conclusions-and-possible-applications}.

Before starting, I would like to stress how this work \emph{is not}
about fitting the contagion curve, or predicting anything; rather,
it is an exercise about the possible theoretical motivations for (\ref{eq:Sornette})
to describe well various cases of epidemic; this does not mean these
reasonings to be completely useless in applicative terms.

\section{Logistic, and generalized logistic, equations\label{sec:Logistic-and-generalized}}

As stated in the § \ref{sec:Introduction}, the ODE (\ref{eq:Sornette})
is a form of \emph{generalized logistic ODE}, because it extends the
original law (\ref{eq:logistica}), to which it reduces for $p=\alpha=1$.
The dynamics (\ref{eq:logistica}) describes a population $C$ that
may grow from arbitrarily small positive amounts up to an equilibrium
value for $C_{\mathrm{eq}}$ so that $\dot{C}\left(C_{\mathrm{eq}}\right)=0$,
that is realized for $C_{\mathrm{eq}}=K$: this parameter $K$ is
referred to as \emph{carrying capacity}.

In order to understand a little bit more the roles of the various
terms in (\ref{eq:logistica}), it is better to re-write it as
\begin{equation}
\dot{C}=rC-\frac{r}{K}C^{2}:\label{eq:ODEC.004.Verhulst}
\end{equation}
in this expression, one may distinguish the competition between an
exponential growth term $rC$ and a self-limitation term $-\frac{r}{K}C^{2}$.
The expression (\ref{eq:ODEC.004.Verhulst}) is of help in understading
the rationale of (\ref{eq:logistica}): as made clear in the expression
(\ref{eq:ODEC.004.Verhulst}), we have a population dynamics in which
all the $C$ individuals take part to both the exponetial growth and
the the self-limitation. The term $rC$ means that, for each of the
$C$ units, one more unit will be ``created'', in a ``reaction''
of the form
\begin{equation}
\mathrm{C}\longrightarrow\mathrm{C}+\mathrm{C},\label{eq:creazione}
\end{equation}
every $\Delta t_{+1}=\frac{1}{r}$ units of time; meanwhile, the term
$-\frac{r}{K}C^{2}$ means that, whenever each of the $C$ individuals
meets another one of the $C$ individuals, one individual is destroyed
in a process
\begin{equation}
\mathrm{C}+\mathrm{C}\longrightarrow\mathrm{C},\label{eq:distruzione}
\end{equation}
that takes place every $\Delta t_{-1}=\frac{K}{rC}$ units of time
(this $\Delta t_{-1}$ becomes smaller and smaller as the amount $C$
increases). Before closing the simple reasoning about (\ref{eq:ODEC.004.Verhulst}),
let me stress that $r$ is the effectiveness of the $+1$ production
process, while $\frac{r}{K}$ measures the effectiveness of the $-1$
destruction process. Expression (\ref{eq:ODEC.004.Verhulst}) is often
referred to as \emph{Verhulst Equation} (VE) \cite{Verhulst}.

It is important to underline the relationship between the mass hypothesis,
and the mathematical way how the two ``reactions'' (\ref{eq:creazione})
and (\ref{eq:distruzione}) are implemented in the kinetics (\ref{eq:ODEC.004.Verhulst})
of the population $C$. Indeed, the fact that the creation rate reads
$rC$ means that \emph{all of the $C$ individuals of the population
do take part to (\ref{eq:creazione})}; in the same way, the expression
$-\frac{r}{K}C^{2}$ for the destruction rate means that \emph{there
is a possible ``annihilation'' for each and every couple of the
$C$ individuals}, being those couples as many as $\mathbb{O}\left(C^{2}\right)$,
because \emph{each} of the $C$ units competes with \emph{all} its
fellows.

Having these considerations about (\ref{eq:logistica}) in mind, we
may re-write the \emph{generalized logistic equation} (\ref{eq:Sornette})
as follows
\begin{equation}
\dot{C}=rC^{p}-\frac{r}{K^{\alpha}}C^{p+\alpha}:\label{eq:Sornette.expanded}
\end{equation}
we now have a production term $rC^{p}$, with $p<1$, and a destruction
(self-competition) term $-\frac{r}{K^{\alpha}}C^{p+\alpha}$, with
$p+\alpha<2$. The interpretation of (\ref{eq:Sornette}) under the
point of view of (fractal) herd behaviour, that is described in §
\ref{sec:Herd-behaviour-and-fractals}, starts from here.

\section{Herd behaviour and fractals\label{sec:Herd-behaviour-and-fractals}}

Let us consider, for instance, two populations $X$ and $Y$, respectively
of preys and predators, living on a surface, e.g. the savnnah, or
a regular portion of the seabed, i.e., 2-dimensional environments.
The predator-prey interaction, consisting of simple predation, gives
rise to a term
\begin{equation}
\dot{X}_{Y}=-kXY\label{eq:preypred.01}
\end{equation}
in a simple Lotka-Volterra model, or to tomething like
\begin{equation}
\dot{X}_{Y}=-\frac{h}{b+X}XY,\label{eq:preypred.02}
\end{equation}
if the model is more sophisticated and a Holling Type II response
function is adopted to describe predation, as in \cite{klepto.01}
($k$ and $h$ are constants). In $\dot{X}_{Y}\left(X,Y\right)$ the
number of preys and predators appears to the first power: in (\ref{eq:preypred.01}),
each of the $X$ preys may ``couple'' with each of the $Y$ predators
with the same destruction rate $k$; in (\ref{eq:preypred.02}), this
happens, but with a rate $\frac{h}{b+X}$ decreasing with the total
amount of preys. Under the idea that \emph{every prey can be reached
by every predator}, there is clearly the mass hypothesis discussed
in § \ref{sec:Introduction}.

In TWT a condition has been introduced \cite{herd.behaviour}, that
changes this hypothesis and, accordingly, modifies the response terms,
the so called \emph{herd behaviour}. Let's suppose that the $X$ preys
move in compact groups of finite size, that cannot be penetrated by
predators: each of the $Y$ predators can only pick preys from the
border of those groups. The right hypothesis is, then, not that each
prey is attacked by each predator, but that just the $X_{\partial}$
preys along the group border will be. So, in the place of $X$ in
$\dot{X}_{Y}\left(X,Y\right)$, one has to put the number of preys
really involved in this predation, i.e. those $X_{\partial}$ ones
sitting on the group border. Because the scene is 2-dimensional, under
the hypothesis of homogeneos surface density (that must be done if
we want to discuss \emph{space-implicit models}, describing everything
via ODEs), comparing the number of individuals along the border of
a geometrical figure with that of all the ones all over the figure
is just as comparing the length of the perimeter with the measure
of the surface. If the figure at hand has a ``size'' $\ell$, clearly
the surface has a measure $A=\mathbb{O}\left(\ell^{2}\right)$, so
that $\ell=\mathbb{O}\left(X^{\frac{1}{2}}\right)$, while the perimeter
scales as $P=\mathbb{O}\left(\ell\right)$: one may conclude
\[
X_{\partial}=\mathbb{O}\left(X^{\frac{1}{2}}\right).
\]
As the preys move just in compact, predator-impenetrable groups, i.e.
as they show herd behaviour, while predators are free to move all
over the 2d space aoutside these groups, the predation terms in (\ref{eq:preypred.01})
and (\ref{eq:preypred.02}) will be re-written as
\begin{equation}
\dot{X}_{Y}=-k'\sqrt{X}Y\ \mathrm{and}\ \dot{X}_{Y}=-\frac{h'\sqrt{X}}{b'+\sqrt{X}}Y\label{eq:preypred.03}
\end{equation}
respectively.

That of the savannah is an $\mathbb{R}^{2}$ example, but the herd
behaviour can be generalized to other geometrical enrivonments: for
instance, if preys and predators move in $\mathbb{R}^{3}$, as it
happens to nekton animals in the sea, then one may state $X=\mathbb{O}\left(\ell^{3}\right)$
and $X_{\partial}=\mathbb{O}\left(\ell^{2}\right)$, so that the predation
terms (\ref{eq:preypred.03}) will read:
\[
\dot{X}_{Y}=-k"X^{\frac{2}{3}}Y,\ \dot{X}_{Y}=-\frac{h"X^{\frac{2}{3}}}{b"+X^{\frac{2}{3}}}Y.
\]
More in general, if those species live in some $\mathbb{R}^{n}$,
but the preys that can be preyed on are segregated in a sub-ambient
$\mathbb{E}_{\mathrm{act}}$ of dimension $\dim\mathbb{E}_{\mathrm{act}}=m\le n$,
clearly some terms as
\begin{equation}
\dot{X}_{Y}=-\hat{k}X^{\eta}Y,\ \dot{X}_{Y}=-\frac{\hat{h}X^{\eta}}{\hat{b}+X^{\eta}}Y,\ /\thinspace\eta=\frac{m}{n}\le1\label{eq:preypred.04}
\end{equation}
will appear in the prey population ODE. Note that, in the expressions
from (\ref{eq:preypred.01}) to (\ref{eq:preypred.04}), a mass hypothesis
is still active on predators, that are supposed to be ``very mobile''
and ``enough mixed'' outside the groups of preys. If, instead, also
the predators are scarcely mobile or slow, possibly packs of predators
interact with herds of preys just via their borders \cite{Collini.et.al.2020},
so that, for instance, one should write
\[
\dot{X}_{Y}=-\tilde{k}\sqrt{XY}\ \mathrm{and}\ \dot{X}_{Y}=-\frac{\tilde{h}\sqrt{XY}}{\tilde{b}+\sqrt{X}}
\]
instead of (\ref{eq:preypred.03}), and so on.

Applying these concepts to the epidemic growth given by equation (\ref{eq:Sornette.expanded})
requires some generalization of the ecological examples just described.
Suppose to deal with a population of infected people $C$ occupying
an environment $\mathbb{E}$ of dimension $\dim\mathbb{E}=\nu$: this
number can be, in principle, any real, positive number, as we are
imaging populations living in any fractal subset of $\mathbb{R}^{n}$.
Suppose that these individuals undergo processes as (\ref{eq:creazione})
and (\ref{eq:distruzione}): if all the individuals living in $\mathbb{E}$
take part to both these processes, clearly the Verhulst equation (\ref{eq:ODEC.004.Verhulst})
will be solved by $C$. Instead, suppose that, in order to be ``active''
pruducing new individuals (when an infected unit meets a susceptible
one), or limiting each other (because when two infected units meet,
no new one appears), those units have to be segregated in a sub-environment
of $\mathbb{E}$, namely some $\mathbb{E}_{\mathrm{act}}\subset\mathbb{E}$,
so that $\dim\mathbb{E}_{\mathrm{act}}=\mu\le\nu$. It is straightforward
to convince ourselves that the active portion of population is
\[
C_{\mathrm{act}}=\mathbb{O}\left(C^{p}\right),\ p=\frac{\mu}{\nu}\le1.
\]
It is then obvious to write the generalization of (\ref{eq:ODEC.004.Verhulst})
to which such a species would undergo:
\begin{equation}
\dot{C}=rC^{p}-\frac{r}{K^{p}}C^{2p},\label{eq:ODEC.005.alpha.eq.p}
\end{equation}
that is precisely the same as (\ref{eq:Sornette.expanded}), or (\ref{eq:Sornette}),
with $\alpha=p$.

Now, what if $\alpha\neq p$ in (\ref{eq:Sornette.expanded})?

Typically, one has $\alpha>p$ as in \cite{Cinesi.Svizzeri}, so it
is sensible to put $\alpha=p+\delta$, with $\delta>0$, and then
re-write (\ref{eq:Sornette.expanded}) as:
\begin{equation}
\dot{C}=rC^{p}-\frac{rC^{\delta}}{K^{p+\delta}}C^{2p}.\label{eq:ODEC.006}
\end{equation}
The only true difference between this case and the ODE (\ref{eq:ODEC.005.alpha.eq.p})
is the fact that the coefficient of $C^{2p}$ depends on $C$ itself.
This can be interpreted in two equivalent ways: on the one hand, one
may say that the effectiveness of the $-1$ limiting process depends
on the population itself as
\begin{equation}
\Delta_{\mathrm{eff}}\left(C\right)\overset{\mathrm{def}}{=}\frac{r}{K^{p+\delta}}C^{\delta},\label{eq:ODEC.008.Delta.eff}
\end{equation}
so that the larger the poulation is, the more destructive the self-limitation
turns out to be among the individuals in $\mathbb{E}_{\mathrm{act}}$;
on the other hand, one might as well state that there is an\emph{
C-local} \emph{effective carrying capacity} $K_{\mathrm{eff}}\left(C\right)$
decreasing with $C$
\begin{equation}
K_{\mathrm{eff}}\left(C\right)\overset{\mathrm{def}}{=}\frac{K^{p+\delta}}{C^{\delta}},\label{eq:ODEC.007.K.eff}
\end{equation}
so that, as the population increases, its dynamics ``sees'' a smaller
and smaller carrying capacity (even if the asymptotic value is still
$C=K$: the only difference with respect to the cases (\ref{eq:logistica})
and (\ref{eq:ODEC.005.alpha.eq.p}) is that in (\ref{eq:Sornette.expanded}),
and hence in (\ref{eq:ODEC.006}), the rush towards the limit $C=K$
gets slower and slower, with respect to the logistic ODE tempo, while
the total population $C$ increases). Another possibile interpretation
of the self-competition term in (\ref{eq:Sornette.expanded}) could
be that, next to the infected people able to infect the others, i.e.
$C_{\mathrm{act}}=\mathbb{O}\left(C^{p}\right)$, there is a class
of infected people with whom those $C_{\mathrm{act}}$ come in contact
uneffectively, that is some $C_{\lim}=\mathbb{O}\left(C^{\alpha}\right)$
\emph{limiting} the contagion, as $-\frac{r}{K^{\alpha}}C^{p}C^{\alpha}$.
In this vision, one should define some geometric locus $\mathbb{E}_{\lim}$,
with Hausdorff dimension $\alpha$, to which the uneffective contacts
are restricted (it is very likely that we well have $\mathbb{E}_{\mathrm{act}}\cap\mathbb{E}_{\lim}\neq\emptyset$).

Now, the crucial point is \emph{to understand why} the COVID-19 contagion
growth, together with other epidemics well studied in the past, should
behave in this way, in terms of the segregation of the various classes
of individuals: a point of view on this, is given in the following
§ \ref{sec:Conclusions-and-possible-applications}.

\section{Conclusions and possible applications\label{sec:Conclusions-and-possible-applications}}

About the interpretation of the powers appearing in (\ref{eq:Sornette}),
for sure one may state that the truely active portion of the infected
people $C_{\mathrm{act}}$ si a very particular function of the whole
number of infected ones, as:
\[
C_{\mathrm{act}}\propto C^{p}.
\]
Similarly, the self-competition term limiting the growth of $C$ according
to (\ref{eq:Sornette}) is a power law in terms of the total of infected
persons, i.e. $\dot{C}_{\lim}\propto C^{p+\alpha}$. The possible
physical interpretation of $p$ and $\alpha$, the one I am suggesting
in this note, is that those non-integer powers should represent \emph{the
geometric locus where ``contagion reactions'' take place}.

In the herd behaviour of TWT, real powers of population variables
represent a measure of the physical places where predators and preys
meet, or where competition takes place, but here such powers must
be attributed a more subtle meaning. While animals in the savannah
move in a 2-dimensional space, see (\ref{eq:preypred.03}), so that
one could state $\dim\mathbb{E}=2$, and $\dim\mathbb{E}_{\mathrm{act}}=1$
(being $\mathbb{E}_{\mathrm{act}}$ the locus where preys can be caught
by predators, the border of prey groups), things are different for
humans infecting each other. The locus $\mathbb{E}$ ``where infected
humans live'' must be understood as a subset of the place where people
live, work and move, i.e., of the network of inhabited centers and
the links connecting them. Let us put $\dim\mathbb{E}=\nu$ (consider
this is far from being easily defined). Moreover, a sub-locus of this
$\mathbb{E}$, i.e. where contagion events really take place, is indicated
as $\mathbb{E}_{\mathrm{act}}$: attributing a value $\dim\mathbb{E}_{\mathrm{act}}=\mu\le\nu$
to the ``dimension'' of $\mathbb{E}_{\mathrm{act}}$ means understanding
which part of the total infected people is really in contact with
susceptible ones, being able to ``produce new infectious people''.
Once this $\mathbb{E}_{\mathrm{act}}$ is identified, its ``size''
should be expressed as a function of the ``size'' of the whole $\mathbb{E}$,
so to be able to write the expression $C_{\mathrm{act}}\left(C\right)$:
provided things work as in the herd behaviour case, i.e. provided
$\mathbb{E}_{\mathrm{act}}$ is a \emph{non-space-filling subset}
of $\mathbb{E}$, hence of Hausdorff dimension smaller than $\dim\mathbb{E}$,
one may state
\begin{equation}
C_{\mathrm{act}}\propto C^{\frac{\dim\mathbb{E}_{\mathrm{act}}}{\dim\mathbb{E}}}=C^{\frac{\mu}{\nu}}\overset{\mathrm{def}}{=}C^{p}.\label{eq:why.p}
\end{equation}
The explanation for the case of $\alpha=p$ is just given by the foregoing
assumption (\ref{eq:why.p}), while, in order to understand the case
$p<\alpha=p+\delta$ one may either think that the ``active'' contagious
people $C_{\mathrm{act}}$ will interact with ``slightly'' more
infected people than themselves alone, so that the limiting locus
will be $\mathbb{E}_{\lim}\supset\mathbb{E}_{\mathrm{act}}\thinspace/\thinspace\dim\mathbb{E}_{\lim}=\alpha\nu$:
this gives rise to the competition term $-\frac{r}{K^{\alpha}}C^{p}\cdot C^{\alpha}$
in the ODE; or think that the coefficient of $C^{2p}$ term in a ``regular''
competition term with a coefficient depending on $C$ explicitly,
as $\dot{C}_{\lim}=-\frac{rC^{\alpha-p}}{K^{\alpha}}C^{2p}$, be this
a competition strength growing as $\mathbb{O}\left(C^{\alpha-p}\right)$,
or a carrying capacity decreasing as $\mathbb{O}\left(C^{p-\alpha}\right)$.

The great question is, then, how to compute $\dim\mathbb{E}$, $\dim\mathbb{E}_{\mathrm{act}}$
and $\dim\mathbb{E}_{\lim}$, provided it makes sense at all to represent
the behaviour of infected humans and of the contagion via fractal
geometrical loci. In this vision, the locus $\mathbb{E}$ should depend
on the human behaviour and society: in particular, it must retrace
the locus where people are concentrated, i.e. the web of the inhabited
centers and communications $\mathcal{H}$. As a fantasy, we could
say $\dim\mathcal{H}=\rho\le2$, since we can at most occupy the 2-dimensional
surface of a country: so $\mathbb{E}\subseteq\mathcal{H}$ will mean
$\nu\le\rho<2$. An indication about $\dim\mathcal{H}$, in agreement
with the arguments here, may be found in \cite{Anversa.frattale},
for example. When one goes from $\dim\mathbb{E}=\nu$ to the value
of $\dim\mathbb{E}_{\mathrm{act}}$, things become more complicated,
because now we have to consider not only the distribution and behaviour
of humans, but also the ``contagion dynamics'', a contribution given
by the nature of the virus. For COVID-19, the contagion seems to take
place via rather close contact, so that particles of the breath of
an infected person are received by the susceptible individual: one
may imagine to select $\mathbb{E}_{\mathrm{act}}$ considering the
sub-locus of $\mathbb{E}$ of the close contacts of the single individual,
i.e. possibly the ``network of personal relationships'' and ``of
casual encounters''. Possibly, this will give $\dim\mathbb{E}_{\mathrm{act}}=\mu\le\nu$,
and hence $p$. Similar considerations will lead to figuring out what
$\alpha$ could be.

\begin{figure}
\begin{centering}
\includegraphics[scale=0.3]{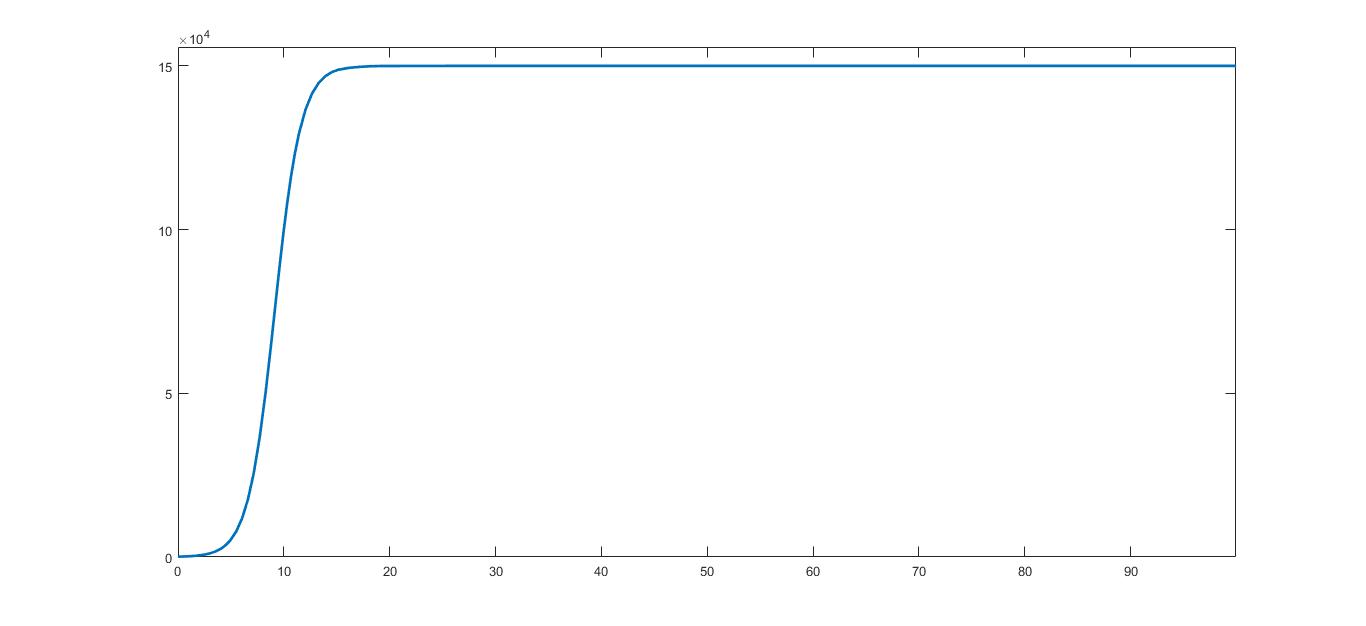}
\par\end{centering}
\caption{\label{figure.01}A curve solving the ODE (\ref{eq:logistica}), i.e.
with $p=\alpha=1$ and $r=0.8\thinspace\mathrm{day}^{-1}$. The initial
value of infected individuals is $C\left(0\right)=100$.}

\end{figure}

\begin{figure}
\begin{centering}
\includegraphics[scale=0.3]{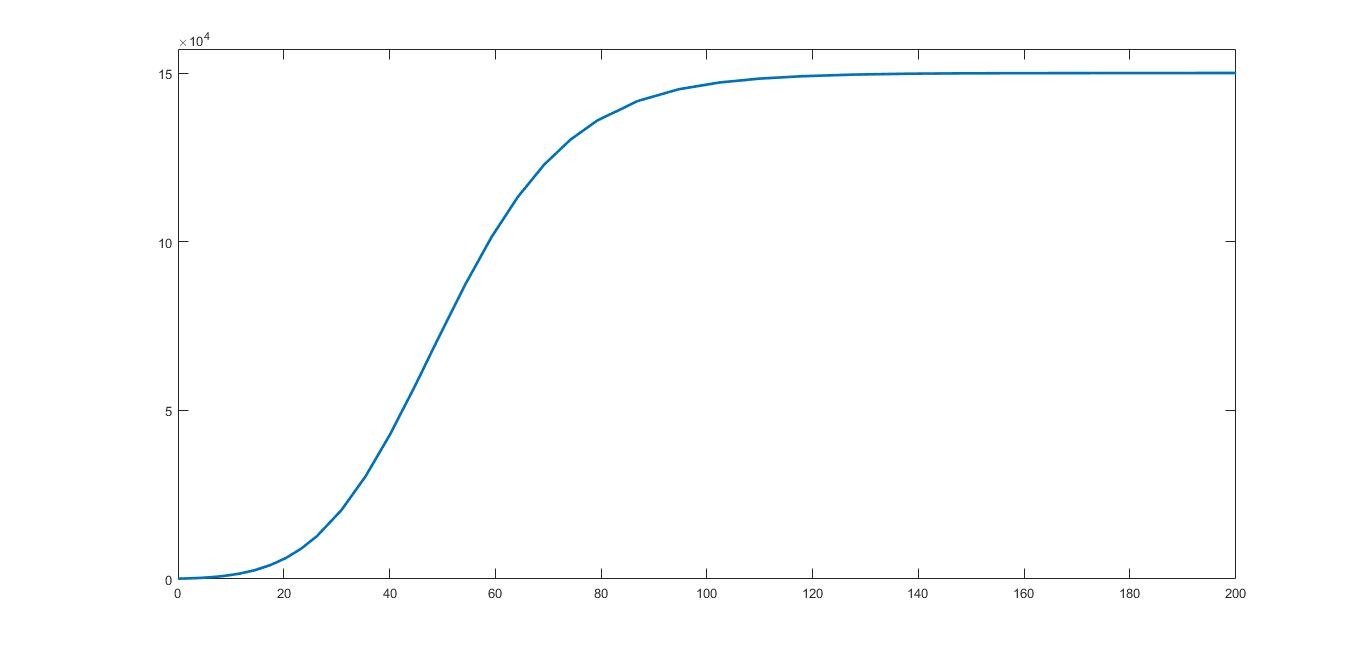}
\par\end{centering}
\caption{\label{figure.02}A curve solving the ODE (\ref{eq:logistica}), i.e.
with $p=0.8$ and $\alpha=1$, and $r=0.8\thinspace\mathrm{people}^{0.2}\cdot\mathrm{day}^{-1}$.
The initial value of infected individuals is $C\left(0\right)=100$.}
\end{figure}

As it is understandable from the aforementioned arguments, ``predicing''
the numbers $\nu$, $\mu$, $p$ and $\alpha$ from what we may study
about the distribution and communications of humans, their relationship
networks, and from what we know about the behaviour of COVID-19, will
be a very tough interdisciplinary task. What one can say by intuition
is that, as the locus $\mathbb{E}_{\mathrm{act}}$ is more sparse,
the behaviour of the outbreak $C\left(t\right)$ will be slower and
slower. For instance: considering $K=15\times10^{4}$ and $r=0.8\thinspace\mathrm{people}^{1-p}\cdot\mathrm{day}^{-1}$,
the curve $C\left(t\right)$ solving the ODE with $p=\alpha=1$ is
the one illustrated in Figure \ref{figure.01}: looking at that plot,
one sees that the maximum value of infected people is reached in practice
between the $20^{\mathrm{th}}$ and the $30^{\mathrm{th}}$ day. If
one puts, instead, $p=0.7$ and $\alpha=1$, the result is that of
Figure \ref{figure.02}: in this case, we see that the value $C\simeq K$
is reached not before $t=140\thinspace\mathrm{days}$, i.e. the growth
is much slower as $p$ decreases.

The slowness of $C\left(t\right)$ towards $K$, increasing with decreasing
$p$, teaches that, with smaller $p$, the outbreak of contagion gives
much more time to the public healthcare administration to take anti-contagion
measures. The faster the reach of $K$ is, the more crowded the hospitals
will be, the more difficult will be to assist ill people, and the
larger the number of dead can be, and this can be mitigated acting
precisely on $p$. Clearly, also acting on $\alpha$ may modify the
shape of the curve $C\left(t\right)$.

Provided equation (\ref{eq:Sornette}) describes COVID-19 outbreak
with time, acting on the exponents in (\ref{eq:Sornette}) may regulate
the time given to a national healthcare administration to confront
it. As argued before, the exponents $p$ and $\alpha$ depend on the
physical distribution of people, on their behaviour and on their relationship
network: under this point of view, one has to hope that the lockdown
meaures taken by many Governments are acting in the direction of diminishing
$p$, and increasing $\delta$ in (\ref{eq:ODEC.008.Delta.eff}),
so to render less and effective the ``creation'' term $rC^{p}$
and more and more rapid the ``cancellation'' term $-\Delta_{\mathrm{eff}}\left(C\right)C^{2p}$.

\end{document}